# Encoding orbital angular momentum of light in space with optical catastrophes

Xiaoyan Zhou[1], John You En Chan[1], Chia-Te Chang[1], Zhenchao Liu[1], Hao Wang [2,3], Andrew Forbes[4], Cheng-Wei Qiu[5, *], Hongtao Wang[1, *], & Joel K. W. Yang[1,6, *]

**Affiliations**
[1]Engineering Product Development, Singapore University of Technology and Design, Singapore 487372, Singapore
[2] Hangzhou International Innovation Institute, Beihang University, Hangzhou 311115, China
[3]School of Instrumentation and Optoelectronic Engineering, Beihang University, Beijing 100191 China.
[4]School of Physics, University of the Witwatersrand, Private Bag 3, Wits 2050, South Africa
[5]Department of Electrical and Computer Engineering, National University of Singapore, Singapore 117583, Singapore
[6]Singapore-HUJ Alliance for Research and Enterprise (SHARE), The Smart Grippers for Soft Robotics (SGSR) Programme, Campus for Research Excellence and Technological Enterprise (CREATE), Singapore 138602, Singapore

**Corresponding author**
*E-mail: chengwei.qiu@nus.edu.sg
*E-mail: hongtao_wang@sutd.edu.sg
*E-mail: joel_yang@sutd.edu.sg

**Abstract**
Light beams carrying orbital angular momentum (OAM) possess an unbounded set of orthogonal modes, offering significant potential for optical communication and security. However, exploiting OAM beams in space has been hindered by the lack of a versatile design toolkit. Here, we demonstrate a strategy to tailor OAM across multiple transverse planes by shaping optical caustics leveraging on catastrophe theory. With complex-amplitude metasurfaces fabricated using two-photon polymerization lithography, we construct these caustics to steer Poynting vectors and achieve arbitrary shapes of OAM beams. Interestingly, we use such an approach to realize "hidden" OAM along the propagation trajectory, where the intensity of the beam is spread out thus avoiding detection. The OAM of these beams can be intrinsic, which avoids OAM distortions arising from the mixing of intrinsic and extrinsic components. By exploiting this intrinsic nature of OAM, we demonstrate the detection of encoded information in optical encryption. Our approach provides a unique framework for dynamic control of OAM in space, with promising applications in optical trapping and sensing, high-capacity data storage, and optical information security.

## Introduction
Among the multiple degrees of freedom (DoFs) of light, its orbital angular momentum (OAM) stands out for its characteristic twisted wavefronts corresponding to an unbounded set of orthogonal modes[1-3]. This property makes OAM an attractive candidate for advancing areas such as high-capacity optical communication[4,5], optical information security[6,7] anti-counterfeiting technology[8] and optical manipulation[9]. Exploiting the selectivity of OAM and its theoretically unbounded spiral mode index enables the multiplexing of numerous OAM-dependent information channels[10-12]. The combination of OAM with other DoFs, such as polarization, wavelength and amplitude, further enhances both the capacity of holographic devices and the security of encryption schemes[6,13,14]. While multiplexing focuses on expanding the number of accessible information channels, the stability of each individual channel depends on the encoding mechanism. To ensure the reliability of OAM-based optical encryption, it is essential to employ the so-called intrinsic OAM, which originates from the internal phase structure of the beam and is independent of any coordinate system[15]. In contrast, extrinsic OAM varies with the external spatial coordinate system. Mixing with extrinsic OAM can introduce OAM distortion and compromise encoding fidelity. Common methods to generate light beams carrying intrinsic OAM include spiral phase plates[16], fork gratings[17,18], mode shaping within microlasers[19,20] and metasurfaces[21]. The gradient method can generate arbitrary OAM patterns at the focal plane, but strong diffraction limits the preservation of their spatial profiles during propagation[22,23]. Exploiting OAM in 3D space provides larger key space for information security, which can be achieved through interference[24-26]. However, the resulting fields are typically restricted to doughnut shapes and exhibit limited flexibility in steering along arbitrary trajectories. Hence, to realize complete spatial control and enhance the information capacity of optical encryption, the ability to manipulate the OAM structures of light across multiple transverse planes is needed.

Here, we propose and demonstrate a strategy to sculpt the OAM structures of light with optical catastrophes. Rooted in singularity theory, catastrophes are abrupt changes in the form of sudden responses of a system to gradual and continuous variations in external conditions[27]. Optical catastrophes manifest as sharp edges known as caustics which form complex light structures, such as the shimmering patterns at the bottom of a swimming pool, and the bright envelopes of light from curved glass surfaces[28-30]. Optical caustics are defined by singularities where a smoothly varying family of rays undergoes abrupt collective focusing into a bright curve[31-33]. This convergence of countless rays in space forms an intensity singularity, which appears in reality a high intensity point (or line), a trivial example of which is the focal spot arising from the convergence of rays of light by a lens. Although caustics are sensitive to perturbation during formation, their resulting caustic structures are robust as they possess topological protection during propagation[27,34,35].

Our strategy to customize the OAM of light on multiple transverse planes is shown in Fig. 1. On each plane, caustic points are arranged along a prescribed path, and a compensation phase[35] is applied to maintain the direction of the local transverse wavevector tangential to the path. Accordingly, energy flow is guided smoothly along the path, yielding prescribed distributions of light intensity, Poynting vector, and OAM density. An example of a caustic beam forming a spiral path is shown in the inset of Fig. 1b. To realize such structured light fields, we fabricate a metasurface using two-photon polymerization lithography (TPL), as shown in Fig. 1a. The metasurface consists of crosslinked polymeric nanofins with varying heights and in-plane orientations that enable complex amplitude modulation. Using this metasurface, we demonstrate a five-dimensional optical information encryption scheme based on spatial position ($x$, $y$, $z$), beam profile, and OAM. This approach provides a versatile design toolkit for sculpting the OAM of light, with potential applications in next-generation optical encryption schemes.

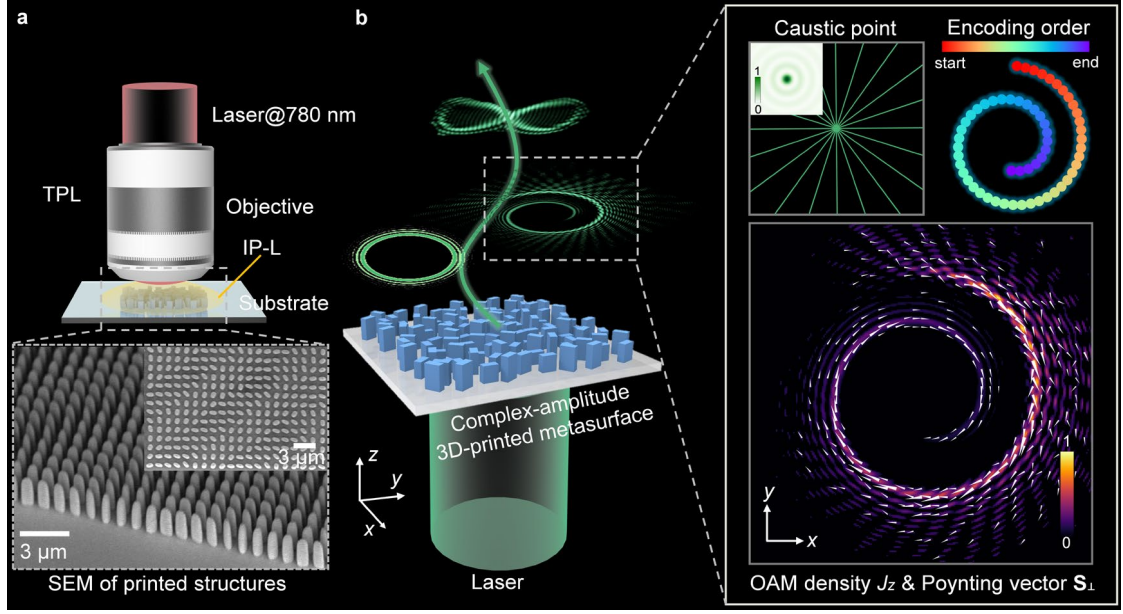

**Fig. 1 Concept of tailoring the OAM of light with caustic points. a** Schematic of the fabrication of 3D-printed nanofin metasurface using TPL. A femtosecond laser source with 780 nm wavelength is focused through an objective lens, and the tightly focused beam induces two-photon polymerization in the IP-L resist. The scanning electron microscopy (SEM) images in the inset show the details of printed nanofins in oblique and top views, respectively. **b** Schematic of an arbitrary patterned caustic beam carrying intrinsic OAM imparted by the 3D-printed metasurface. The inset displays a spiral structure of OAM and Poynting vector, with caustic points tracing the spiral path. The background color represents the normalized OAM density, while arrows indicate the Poynting vector.

## Results
### Guiding OAM of light with caustic points

Given that the field is defined with zero spin angular momentum, the total OAM density at any plane along the $z$ axis is expressed as:

$$J_z = \iint_R \left( \mathbf{r} \times \mathbf{S} / c^2 \right)_z d^2 \mathbf{r}. \tag{1}$$

Here, $\mathbf{r} = (x, y, z)$ is the spatial position in the Cartesian coordinate system. $\mathbf{S} = \varepsilon_0 c^2 \langle \mathbf{E} \times \mathbf{B} \rangle$ is the time averaged Poynting vector, in which $\varepsilon_0$ is the free space permittivity, and $c$ is the speed of light in vacuum, $\mathbf{E}$ and $\mathbf{B}$ are the electric and magnetic fields, respectively. The subscript $R$ denotes the region of integration. Consequently, the average OAM is given by

$$\langle l \rangle = \omega c J_z / \iint_R S_z d^2 \mathbf{r} \tag{2}$$

where $\omega$ is the angular frequency and $S_z$ is the z-component of $\mathbf{S}$. It is a real number, and $\langle l \rangle \hbar$ represents the mean OAM per photon or expectation value of OAM with $\hbar$ being reduced Planck constant.

We first consider controlling the OAM of light in the 2D case, where the OAM remains invariant across all transverse planes. Detailed derivation is shown in Fig. 2a. For any propagation-invariant beam, its transverse wavevector is restricted to a ring and the electric field in the far field can be described by Whittaker's integral[36]

$$E(\mathbf{r}) = \int_{-\pi}^{\pi} F(\varphi) \exp\left[ik_\perp \mathbf{r}_\perp \cdot \mathbf{u}(\varphi)\right] d\varphi. \tag{3}$$

Here, $\varphi$ denotes the azimuthal angle of the transverse wavevector in the Fourier plane, and the integration is carried out over the full angular range $[-\pi, \pi]$. $k_\perp = \sqrt{k_x^2 + k_y^2}$ is magnitude of the transverse wavevector; $\mathbf{r}_\perp = (x, y)$ is the transverse position vector; $\mathbf{u}(\varphi) = (\cos\varphi, \sin\varphi)$ is the unit vector; $F(\varphi) = A(\varphi)\exp[i\Phi(\varphi)]$ is the angular spectrum of the light field with the amplitude $A(\varphi)$ and phase $\Phi(\varphi)$ in Fourier space. According to the principle of catastrophe optics, caustics are obtained by applying the stationary phase method to Eq. (3), yielding $\mathbf{r}_c(\varphi) = [\Phi''(\varphi)\mathbf{u}(\varphi) - \Phi'(\varphi)\mathbf{u}'(\varphi)]/k_\perp$. Note that $\Phi'(\varphi)$ and $\Phi''(\varphi)$ are the first and second derivatives of $\Phi(\varphi)$. To generate a cusp caustic at $\mathbf{r}_c(\varphi) = (x_0, y_0)$ as a tool for guiding the OAM of light, we set $\mathbf{r}_c'(\varphi) = 0$ to obtain a point phase $\Phi(\varphi) = -k_\perp \mathbf{r}_c \cdot \mathbf{u}(\varphi)$. The amplitude is set as $A(\varphi) = 1$ for simplicity.

To realize a specific OAM structure, we arrange these caustic points along a transverse path $\mathbf{r}_p(\tau)$ with $\tau$ being arc length of a parametric curve. The point phase on the path is expressed as a path phase $-k_\perp \mathbf{r}_p \cdot \mathbf{u}(\varphi)$. In order to guide the energy flow along this path, we need to introduce a compensation phase $\phi_c(\tau) = k_\perp \tau$ to ensure that the transverse wavevectors remain tangential to this path and to maximize the light intensity along it. Thus, the angular spectrum is given by $F(\varphi) = \int_0^T \exp\{ik_\perp[\tau - \mathbf{r}_p(\tau) \cdot \mathbf{u}(\varphi)]\} d\tau$ with $T$ being the end of a transverse path. The optical field in real space can be obtained by performing a Fourier transform of the obtained angular spectrum, written as

$$E(\mathbf{r}) = \int_0^T \exp(ik_\perp \tau) J_0\left(k_\perp \left|\mathbf{r}_\perp - \mathbf{r}_p(\tau)\right|\right) d\tau. \tag{4}$$

We can clearly observe that the electric field is a superposition of zero-order Bessel functions $J_0(\cdot)$ centered along the path $\mathbf{r}_p(\tau)$.

To demonstrate how the OAM of light is controlled by designing the caustic points, we present the cases of circular and astroid caustics as examples, as shown in Fig. 2b and Fig. 2c, respectively. Initially, the caustic points on the cross-section are arranged along the prescribed paths and sequentially superposed along the path indicated by the rainbow colors, as shown in Fig. 2b(i) and Fig. 2c(i). To guide the energy flow along the preset path, a gradient phase which grows with the arc length of paths is required, as illustrated in Fig. 2b(ii) and Fig. 2c(ii). The resulting angular spectra for these two cases are displayed in Fig. 2b(iii) and Fig. 2c(iii). The distribution of the family of rays is determined using the stationary phase method, where the envelope of these rays is either a circular or an astroid caustic, as depicted in Fig. 2b(iv) and Fig. 2c(iv). These caustic structures define the boundaries of discontinuity in the ray density and define the regions of peak intensity in the light field, as shown in Fig. 2b(v) and Fig. 2c(v). The corresponding Poynting vectors, represented by red arrows in Fig. 2b(vi) and Fig. 2c(vi), indicate both the direction and magnitude of energy flow in the transverse plane. Notably, the energy flow aligns with the superposition direction of the caustic points. Thus, the sign of the average OAM can be reversed by changing this direction. Finally, as shown in Fig. 2b(vii) and Fig. 2c(vii), the OAM density profile conforms with the pre-designed circular or astroid path, hence demonstrating the successful construction of the OAM structures using caustic points. Due to the axial symmetry of the field distribution, the intensity centroid is located near the coordinate origin, and the total momentum is primarily directed along the propagation axis. Under these conditions, the contribution of extrinsic OAM can be neglected. Consequently, the OAM arises almost entirely from the intrinsic phase structure of the field, and the intrinsic component accounting for as much as 99.99%. The calculation method is detailed in Supplementary Note 1. This intrinsic OAM is invariant to beam misalignment and propagation-induced shifts, thus ensuring accurate retrieval of the encoded OAM, suited to optical encryption applications.

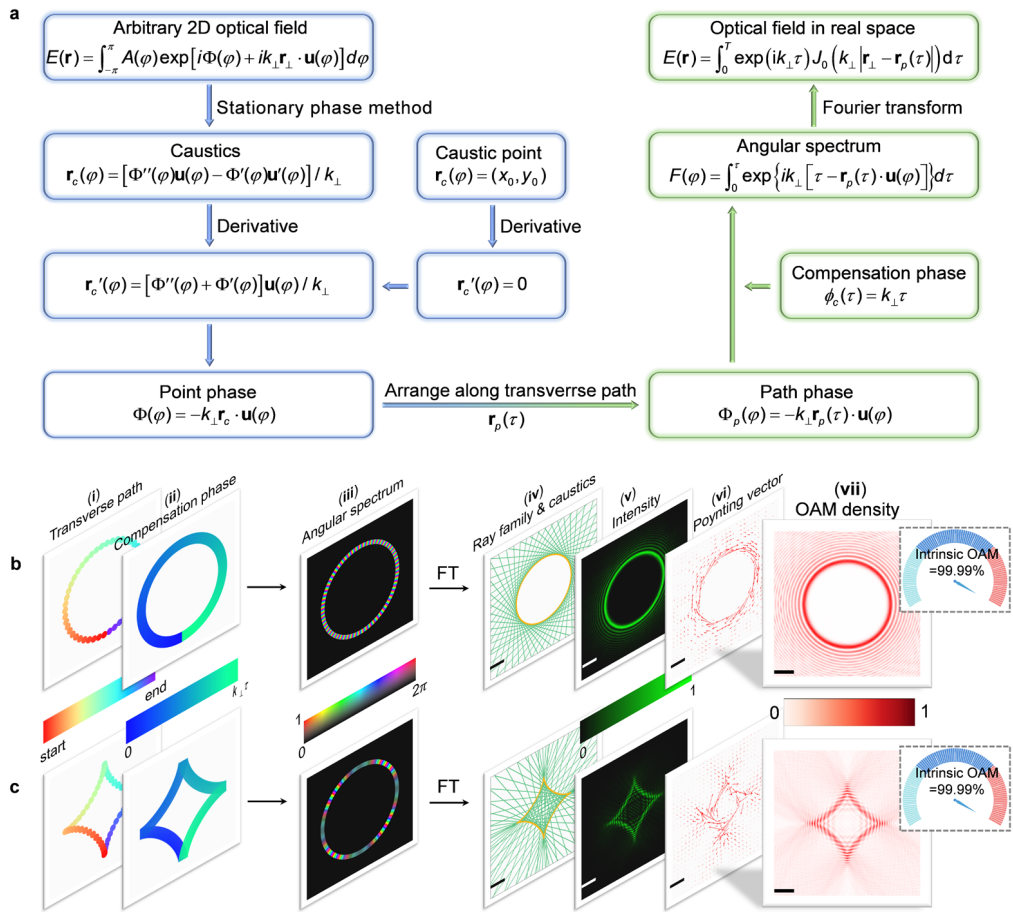

**Fig. 2 Engineering the OAM of light with caustic points in the 2D case. a** Derivation map for constructing propagation-invariant OAM beams consisting of caustic points, e.g. **b** Circular case, and **c** Astroid case. Calculations showing (**i**) the superposition order of caustic points, (**ii**) corresponding compensatory phase, (**iii**) the angular spectrum in Fourier space, (**iv**) ray families and their envelope, (**v**) intensity, (**vi**) Poynting vector, and (**vii**) OAM density distributions in real space. The inset in the upper-right corner shows the proportion of intrinsic OAM. FT, Fourier transform. Scale bars, $50\lambda$. The average OAM value $\langle l \rangle$ is confirmed by performing modal

analysis (Supplementary Note 2).

**Customizing OAM variants of light**

We further extend the approach to achieve arbitrary profile of OAM beams in the 3D case using analytical equations (details in Supplementary Note 3). Two representative beam variants are demonstrated: one exhibiting a sinusoidal propagation trajectory, and another apparently disappears at an intermediate propagation distance.

To demonstrate the first concept, we tailor the OAM of light into an astroid shape along a sinusoidal curved propagation trajectory. The trajectory is denoted as $X(z) = 0$, $Y(z) = 100\lambda \sin z$, lying on the plane $x = 0$. The maximum propagation distance is set as $z_{max} = 5 \times 10^4 \lambda$. For clarity, we select three specific planes $z_1 = 0.15 z_{max}$, $z_2 = 0.5 z_{max}$, and $z_3 = 0.85 z_{max}$ along the trajectory, to illustrate the formation of the predefined OAM structures in the transverse plane with caustic points. Across these three selected planes, the caustic points are arranged along an astroid path and sequentially superposed following the rainbow colors. To ensure the transverse wavevectors remain tangential to the path and the constructive interference occurs at every caustic point, the compensation phase, $\phi_{length}(\tau) + \phi_{trajectory}(\tau)$, is applied throughout the process. As depicted in the first branch of Fig. 3a, when only $\phi_{length}(\tau)$ is considered, the resulting caustics field does not show as intended design. This deviation arises from the propagation trajectory, which necessitates a rigorous phase compensation to accurately construct the desired field distribution. The second branch of Fig. 3a shows the results after applying the full compensation phase $\phi_{length}(\tau) + \phi_{trajectory}(\tau)$, where the angular spectrum contains both correct amplitude and phase information. The Fourier transform of this angular spectrum produces the desired optical field in real space, as shown in Figs. 3b-d. In addition, the stationary phase method is employed to visualize the caustic surface propagating along the sinusoidal trajectory and the projection of rays onto the three specific planes, as shown in the right panel of the second branch in Fig. 3a. The envelope of these ray families precisely defines the caustic structure. Figures 3b-d present the distributions of intensity, Poynting vector, and OAM density on these three planes. During the propagation, both the intensity and the OAM density preserve their astroid profiles, while the Poynting vector exhibits an astroid-shaped energy flow, as shown with the red arrows. The average OAM is 3 on each transverse plane, with the intrinsic component contributing 98.58%~99.88% (Supplementary Note 1).

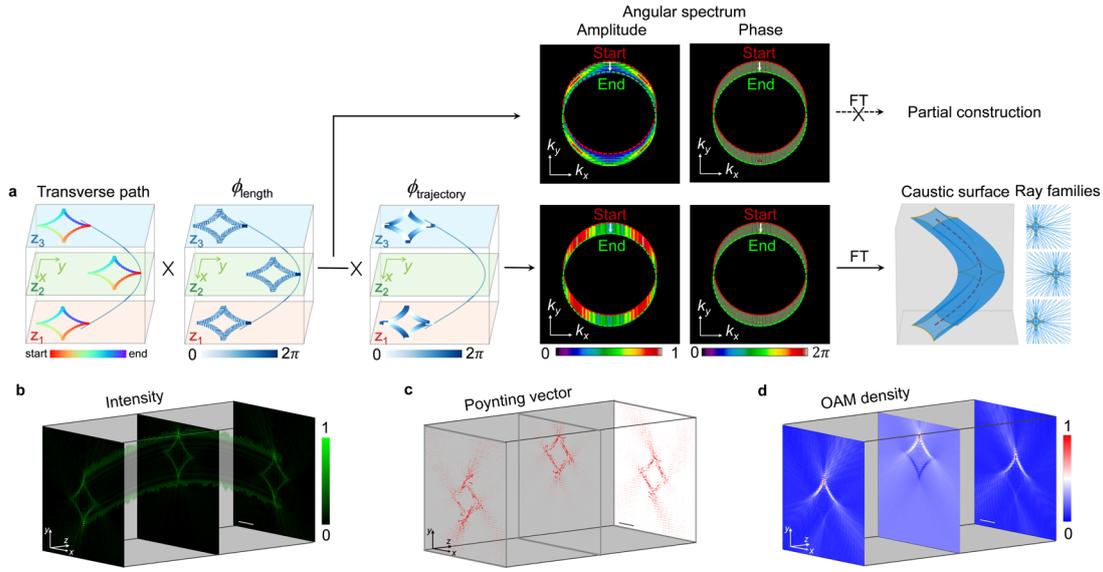

**Fig. 3 Customizing OAM of light with caustic points in the 3D case. a** The effect of partial and full compensation phase on shaping astroid OAM along a sinusoidal propagation trajectory. **b-d** Intensity profile, Poynting vector and OAM density of the beams with full compensation phase. FT, Fourier transform. Scale bars, $100\lambda$.

**Hidden OAM**

To establish an unprecedented optical encryption approach resistant to eavesdropping and unauthorized detection, we propose a new OAM variant termed hidden OAM. Its intensity is spread out over a broad spatial region during propagation to conceal the beam. In our strategy, caustics at a fixed transverse plane in real space correspond to a complex-amplitude ring in Fourier space. By superposing these rings, caustics at different propagation planes can be tailored independently, enabling the dynamic evolution of OAM structures of light. Furthermore, the hidden OAM can surprisingly be achieved with some missing complex-amplitude rings.

As an example, we present the hiding of OAM beams wherein the OAM density initially exhibits an astroid distribution, disappears, and then re-emerges as a five-cusped morphology. The underlying concept of hidden OAM is illustrated in Fig. 4a. We still choose the three propagation distances as above to visualize the evolution of fields. The superposition sequence of caustic points follows the rainbow colors, with compensation phase corrections applied accordingly, as shown in Fig. 4b. In this design, the propagation trajectory is a straight line. The angular spectral response is shown in Fig. 4c. The absence of caustic points in the intermediate region results in zero complex amplitude. By applying the Fourier transform, we construct the corresponding optical field distribution in real space. The intensity profile, Poynting vector, and OAM density distribution derived from this design are displayed in Fig. 4d. Throughout the process, the OAM of light can be steered into a particular spatial profile, and then it vanishes where energy disperses across the transverse plane, and the relative light intensity of these speckles is reduced by two orders of magnitude. Subsequently, the energy and momentum redistribute and converge to a different profile. Notably, the empty form does not signify the physical disappearance of OAM. Instead, it arises from the diffusion of energy and phase gradients into a broader spatial domain, and the average OAM value and energy are conserved. Nevertheless, the global phase information of the OAM is preserved throughout the process and can be revealed under ingeniously designed conditions. Leveraging on this property, the empty form can serve as a concealed encoding state, effectively hiding structural information during transmission, thereby substantially increasing the complexity of information concealment and enhancing resistance to eavesdropping and brute-force attacks.

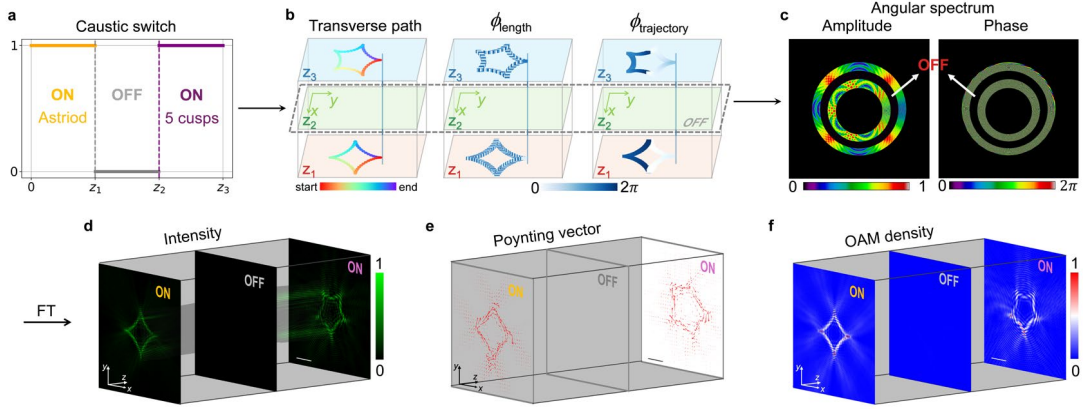

**Fig. 4 OAM switching from astroid to five-cusped structures produced by optical caustics. a** Conceptual design of a caustic switch, where the binary value 1 represents the ON state and 0 represents the OFF state. **b** Superposition order and corresponding compensation phase. **c** Complex-amplitude distributions in Fourier space. **d-f** Intensity profile, Poynting vector, and OAM density distribution in real space. Scale bars, $100\lambda$.

**OAM Match integrated with 3D-printed metasurfaces**

To further verify the OAM encoding and its application in high-dimensional information encryption, we fabricate a 3D-printed metasurface composed of a 6×6 OAM beam array using TPL (see Methods and Supplementary Movie for design and fabrication details). Figure 5a presents a photograph of the fabricated sample together with SEM images in top and oblique views. Figure 5b illustrates the design of the OAM beam array. The design target (first column) is the combination of two key components: the intensity profiles of OAM beams (second column) and average OAM (third column), which are visually illustrated by different background colors. The implementation details for each OAM unit are provided in Supplementary Note 4. The system is designed to sequentially match and eliminate blocks with identical structures along the propagation trajectory, analogous to the rule of the game Candy Crush. A lookup table correlating these six caustic beam shapes with their corresponding average OAM is shown in Fig. 5d.

    We observe the optical performance of the light fields using a custom-built optical setup (see Methods for characterization details). The results for the fabricated sample are shown in Fig. 5c. At propagation distance $z_1$, two "∞" patterns with an average OAM of zero are joined by the white dashed line to show the elimination of OAM beams, that disappear at $z_2$. Meanwhile, two "○" patterns with an average OAM of eight are matched and eliminated at $z_3$. The average OAM is determined by measuring the pattern size of the experimental patterns and retrieving the corresponding values from a database of simulated results, as shown in Fig. 5d. We introduce a hexadecimal four-digit ID (i.e., 12A6) to identify each OAM unit in the array. The first two digits represent the spatial coordinates, with the first digit to fourth digits denoting the row, column, the pattern profile and average OAM, respectively. For example, the code "12A6" refers to a pattern located at row 1, column 2, with geometry "A" and an average OAM of six. Based on this

decoding rule, all four-digit hexadecimal numbers at the three propagation distances can be decrypted through four-digit IDs. A customized program converts character groups into ASCII strings using the Base64 algorithm to output the username, password, and additional identification code for login. This decryption process is illustrated in Fig. 5e. The multi-plane strategy enables information encoding in space. Another example of a Tetris-like OAM beam array is provided in Supplementary Note 5. Our proposed scheme serves as a platform for five-dimensional optical encryption (spatial position ($x$, $y$, $z$), beam profile, and OAM) with an achievable information capacity of $I = \log_2 A \times B \times M \times P^K$ bits, where $A \times B$ denotes the number of beam units, $M$ the average OAM, $P$ the number of beam-profile classes, and $K$ the number of observable transverse planes. While these degrees of freedom are unbounded in theory, practical fabrication and detection constraints restrict the number of distinguishable states.

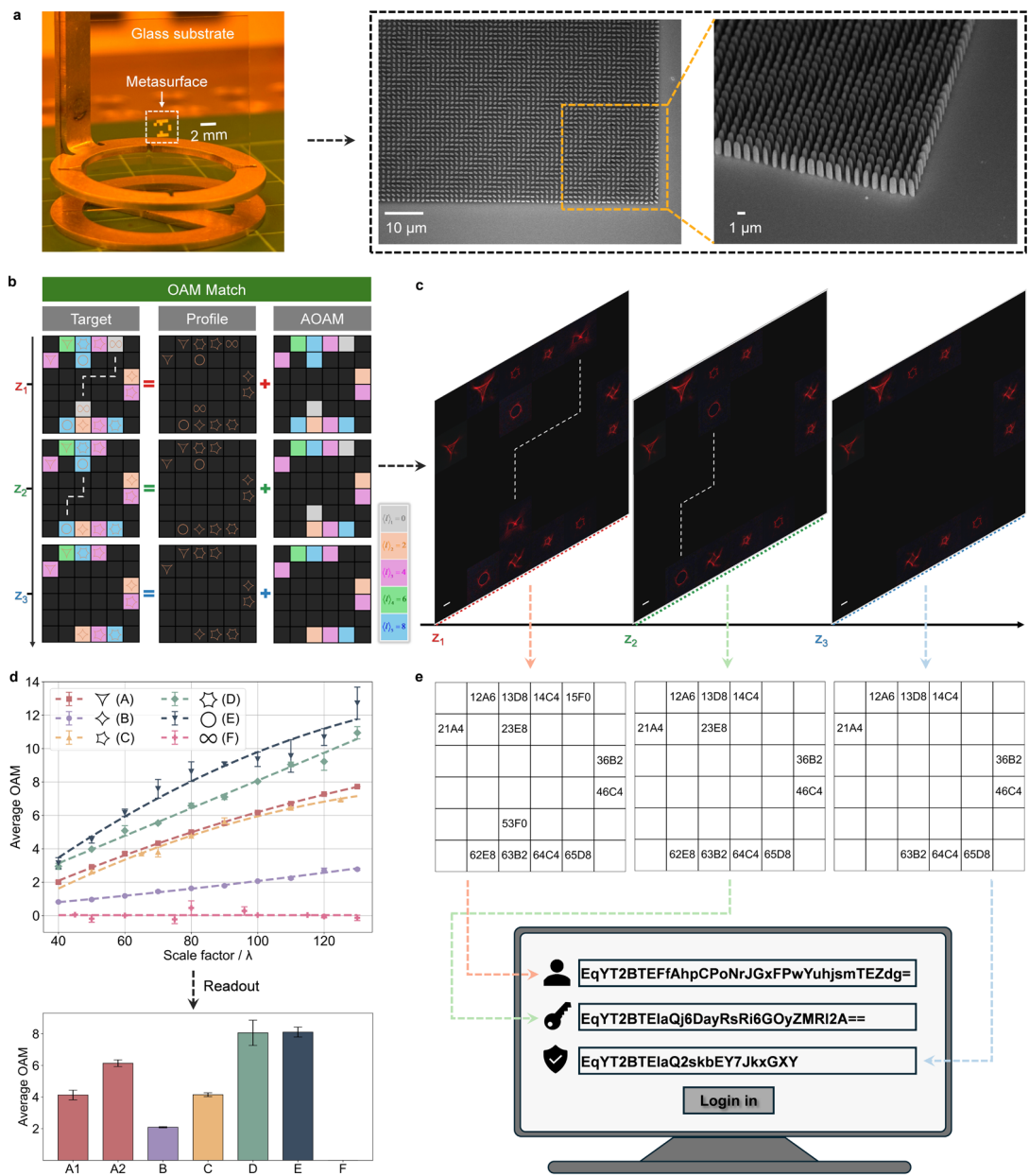

**Fig. 5 Proof-of-concept experimental demonstration of optical encryption. a** Photograph of the metasurface together with magnified SEM images in top and oblique views. **b** Schematic illustration of a 6 × 6 OAM beam array. The first column shows the target OAM beams at propagation distances $z_1$, $z_2$, and $z_3$. OAM density profiles in the second column and the average OAM in the third column are tailored as design elements. **c** Measured transverse intensity patterns at three observation planes. **d** Averaged OAM values are plotted as a function of scale factor and read out from the plot based on experimental results. **e** Decryption of the hexadecimal number combination, revealing plaintext messages including the username, password, and identification code.

**Discussion**

We demonstrated that the OAM of light can be sculpted into arbitrary patterns in space with caustic points driven by optical catastrophes. By applying a compensation phase at the caustic points, the transverse wavevectors remain tangential to the path, and the optical intensity is concentrated along it, enabling flexible encoding of OAM structures. The average OAM value is associated with geometric scaling of intensity pattern, and arbitrary shaped pattern can be realized for the same average OAM value. Shifting the center of the angular-spectrum ring further extends this capability to generate distinct OAM structures across multiple planes and even hidden OAM, wherein the OAM structure can be concealed and reconstructed on demand during propagation. At the same time, the approach enables the generation of OAM beams with high intrinsic content while reducing distortions in optical encoding caused by the mixing of intrinsic and extrinsic components. As proof of concept, we experimentally demonstrated a five-dimensional (spatial position (*x*, *y*, *z*), beam profile, and OAM) optical information encryption scheme with a complex-amplitude 3D-printed metasurface. Benefiting from the fast prototyping and cost-effectiveness of TPL, these metasurfaces with high resolution, an ultra-thin profile, and sub-millimeter extension can be flexibly fabricated to facilitate integration into photonic systems. The information encoded on the metasurface across multiple propagation planes, in which each plane comprises a 6×6 OAM beam array. This multidimensionality significantly increases the complexity of the encryption scheme and enhances its security against unauthorized access. The average OAM is obtained directly from the experimental scale factor in current work, but could be measured more accurately by analyzing the OAM spectrum in future work[37,38]. Looking ahead, introducing OAM into caustic beams provides a versatile platform for high-dimensional, robust, and programmable control of structured light in next-generation optical communication, data storage, information encryption, and optical manipulation.

**Methods**

**Design and fabrication of a 3D-printed metasurface**

The metasurface consists of nanofins made from polymerized IP-L photoresist (refractive index~1.52 in the visible range). The nanofins exhibit uniform transverse dimensions (width = 400 nm, length = 800 nm), whereas their heights vary from 3.0 μm to 3.7 μm (see Supplementary Note 6 for design details). The height and in-plane rotation angle of each nanofin serve as independent parameters to modulate the amplitude and phase of the transmitted cross-polarized light, respectively. Each OAM beam unit of the metasurface contains an array of 300 × 300 nanofins with a periodicity of 1.25 μm, resulting in a unit area of 375 μm × 375 μm.

This metasurface is fabricated using Nanoscribe Photonic Professional GT system. All structures are written on indium tin oxide-coated glass substrates using a ×63 Plan-Apochromat objective lens (NA = 1.40) in a dip-in configuration with IP-L photoresist (refractive index ~1.52 in the visible range). The fabrication process operates in continuous mode under galvo scan mode, with a scan speed of 7000 μm/s, laser power of 45 mW, slicing step size of 20 nm, galvo settling time of 2 ms, and piezo settling time of 20 ms. Each write area is 100 μm × 100 μm unit cell. After exposure, the samples undergo a multi-step development process: they are first immersed in propylene glycol monomethyl ether acetate for 15 minutes, then transferred to isopropyl alcohol for 5 minutes (simultaneously exposed by a Dymax BlueWaveMX-150UV LED curing system set at 70% maximum power), and immersed in methoxynonafluorobutane for 7 minutes. Finally, the samples are dried in ambient air through evaporation.

**Optical set-up for characterization**

The optical setup is shown in Supplementary Note 7. A 632 nm laser beam from an NKT supercontinuum laser is converted to be circularly polarized by a linear polarizer and a quarter-wave plate and projected onto the sample. The transmitted light is magnified using a microscope objective (Nikon, NA = 0.45, 20×) and a tube lens ($f = 200$ mm). The quarter-wave plate and polarizer only allow the cross-polarized beam to transmit. Finally, the output is collected by a high-resolution CMOS camera (DCC3260C, Thorlabs).

**Data availability**

All data supporting the findings of this study are documented within manuscript and Supplementary Information. The data is available from the corresponding authors on request.

**References**

1. Allen, L., Beijersbergen, M.W., Spreeuw, R. & Woerdman, J. Orbital angular momentum of light and the transformation of Laguerre-Gaussian laser modes. *Physical review A* **45**, 8185 (1992).
2. Simpson, N., Dholakia, K., Allen, L. & Padgett, M. Mechanical equivalence of spin and orbital angular momentum of light: an optical spanner. *Optics letters* **22**, 52-54 (1997).
3. Leach, J., Padgett, M.J., Barnett, S.M., Franke-Arnold, S. & Courtial, J. Measuring the orbital angular momentum of a single photon. *Physical review letters* **88**, 257901 (2002).
4. Yan, Y. *et al.* High-capacity millimetre-wave communications with orbital angular momentum multiplexing. *Nature communications* **5**, 4876 (2014).
5. Lei, T. *et al.* Massive individual orbital angular momentum channels for multiplexing enabled by Dammann gratings. *Light: Science & Applications* **4**, e257-e257 (2015).
6. Ouyang, X. *et al.* Synthetic helical dichroism for six-dimensional optical orbital angular momentum multiplexing. *Nature photonics* **15**, 901-907 (2021).
7. Liu, S. *et al.* Optical encryption in the photonic orbital angular momentum dimension via direct-laser-writing 3D chiral metahelices. *Nano Letters* **23**, 2304-2311 (2023).
8. Wang, H. *et al.* Coloured vortex beams with incoherent white light illumination. *Nature Nanotechnology* **18**, 264-272 (2023).
9. Molina-Terriza, G., Torres, J.P. & Torner, L. Twisted photons. *Nature physics* **3**, 305-310 (2007).

10. Ren, H. *et al.* Complex-amplitude metasurface-based orbital angular momentum holography in momentum space. *Nature Nanotechnology* **15**, 948-955 (2020).
11. Ren, H. *et al.* Metasurface orbital angular momentum holography. *Nature Communications* **10**, 2986 (2019).
12. Fang, X., Ren, H. & Gu, M. Orbital angular momentum holography for high-security encryption. *Nature Photonics* **14**, 102-108 (2020).
13. Yang, H. *et al.* Angular momentum holography via a minimalist metasurface for optical nested encryption. *Light: Science & Applications* **12**, 79 (2023).
14. Zhou, H. *et al.* Polarization-Encrypted Orbital Angular Momentum Multiplexed Metasurface Holography. *ACS Nano* **14**, 5553-5559 (2020).
15. O'neil, A., MacVicar, I., Allen, L. & Padgett, M. Intrinsic and extrinsic nature of the orbital angular momentum of a light beam. *Physical review letters* **88**, 053601 (2002).
16. Beijersbergen, M., Coerwinkel, R., Kristensen, M. & Woerdman, J. Helical-wavefront laser beams produced with a spiral phaseplate. *Optics communications* **112**, 321-327 (1994).
17. Genevet, P., Lin, J., Kats, M.A. & Capasso, F. Holographic detection of the orbital angular momentum of light with plasmonic photodiodes. *Nature communications* **3**, 1278 (2012).
18. Huang, K. *et al.* Spiniform phase-encoded metagratings entangling arbitrary rational-order orbital angular momentum. *Light: Science & Applications* **7**, 17156-17156 (2018).
19. Huang, C. *et al.* Ultrafast control of vortex microlasers. *Science* **367**, 1018-1021 (2020).
20. Forbes, A., Mkhumbuza, L. & Feng, L. Orbital angular momentum lasers. *Nature Reviews Physics* **6**, 352-364 (2024).
21. Li, R. *et al.* Broadband Continuous Integer- and Fractional-Order Multimode OAM Beam Generator via a Metasurface. *ACS Photonics* **12**, 870-878 (2025).
22. Zhang, X. *et al.* Multiplexed generation of generalized vortex beams with on-demand intensity profiles based on metasurfaces. *Laser & Photonics Reviews* **16**, 2100451 (2022).
23. Zhang, X. *et al.* Basis function approach for diffractive pattern generation with Dammann vortex metasurfaces. *Science Advances* **8**, eabp8073 (2022).
24. Dorrah, A.H., Rubin, N.A., Tamagnone, M., Zaidi, A. & Capasso, F. Structuring total angular momentum of light along the propagation direction with polarization-controlled meta-optics. *Nature communications* **12**, 6249 (2021).
25. Yang, J. *et al.* Transformation of longitudinally customizable curved vector vortex beams using dielectric metasurface. *Laser & Photonics Reviews* **18**, 2400226 (2024).
26. Dorrah, A.H., Palmieri, A., Li, L. & Capasso, F. Rotatum of light. *Science Advances* **11**, eadr9092 (2025).
27. Arnold, V.I., Wassermann, G. & Thomas, R. *Catastrophe theory*, (Springer, 1986).
28. Nye, J.F. *Natural focusing and fine structure of light: caustics and wave dislocations*, (CRC Press, 1999).
29. Berry, M. *A Half-Century of Physical Asymptotics and Other Diversions: Selected Works by Michael Berry*, (World Scientific, 2017).
30. Born, M. & Wolf, E. *Principles of optics: electromagnetic theory of propagation, interference and diffraction of light*, (Elsevier, 2013).
31. Kravtsov, Y.A. & Orlov, Y.I. *Caustics, catastrophes and wave fields*, (Springer Science & Business Media, 2012).
32. Poston, T. & Stewart, I. *Catastrophe theory and its applications*, (Courier Corporation, 2014).

33. Berry, M.V. & Upstill, C. IV catastrophe optics: morphologies of caustics and their diffraction patterns. *Progress in optics* **18**, 257-346 (1980).
34. Spaegele, C.M. *et al.* Topologically protected optical polarization singularities in four-dimensional space. *Science Advances* **9**, eadh0369 (2023).
35. Zhou, X. *et al.* Arbitrary engineering of spatial caustics with 3D-printed metasurfaces. *Nature Communications* **15**, 3719 (2024).
36. Whittaker, E.T. On the partial differential equations of mathematical physics. *Mathematische Annalen* **57**, 333-355 (1903).
37. D'Errico, A., D'Amelio, R., Piccirillo, B., Cardano, F. & Marrucci, L. Measuring the complex orbital angular momentum spectrum and spatial mode decomposition of structured light beams. *Optica* **4**, 1350-1357 (2017).
38. Wang, Z. *et al.* Universal analyzer for measuring the orbital angular momentum spectrum of a randomly fluctuated beam. *Optics Letters* **49**, 7250-7253 (2024).

**Acknowledgements**

J.K.W.Y. acknowledges funding support from the National Research Foundation of Singapore (NRFS), under its Competitive Research Programme award NRF-CRP30-2023-0003 and NRF Investigatorship Award NRF-NRFI06-2020-0005. C.-W.Q. acknowledges the financial support of the National Research Foundation, Prime Minister's Office, Singapore under Competitive Research Program Award NRF-CRP22-2019-0006. H.W. acknowledges the support from the Research Funding of Hangzhou International Innovation Institute, Beihang University (Grant No. 015733207-000001) and the Fundamental Research Funds for the Central Universities (Grant No. 501RCQD2025117002).

**Author contributions**

X.Y.Z. conceived the idea and performed the design with assistance from H.T.W. and C.-W.Q. X.Y.Z. conducted numerical simulation. X.Y.Z. executed fabrication, and optical characterization with the help of H.T.W. C.T.C. carried out the SEM characterization. X.Y.Z. prepared figures and drafted the manuscript with assistance from H.T.W and J.Y.E.C. All the authors contributed to the data analysis and manuscript revision. J. K.W.Y. and C.-W.Q. supervised the whole project.

**Competing interests**

The authors declare no competing interests.